\documentclass[format=acmsmall, review=false, screen=true]{acmart}

\usepackage{booktabs} 
\usepackage{amsmath}
\usepackage{mathtools}
\usepackage{dsfont}
\usepackage{tikz}
\usepackage{pst-3d,tikz-3dplot,multido}

\usepackage{tikz}
\usepackage{psfrag}
\usepackage{graphics}
\usepackage{breakurl}

\definecolor{matlab_blue}{rgb}{0,0.4470,0.7410}
\definecolor{matlab_red}{rgb}{0.8500,0.3250,0.0980}

\usetikzlibrary{shapes}
\usetikzlibrary{patterns}
\usetikzlibrary{arrows, decorations.pathmorphing,snakes}

\DeclareMathOperator{\lcm}{lcm}
\DeclareMathOperator{\diag}{diag}
\DeclareMathOperator{\argmin}{arg\,min}
\DeclarePairedDelimiter\ceil{\lceil}{\rceil}
\DeclarePairedDelimiter\floor{\lfloor}{\rfloor}
\let\mod\relax\DeclareMathOperator*{\mod}{mod}

\usepackage{algorithm,algpseudocode}
\algblock{Inputs}{EndInputs}
\algnotext{EndInputs}
\algblock{Outputs}{EndOutputs}
\algnotext{EndOutputs}
\algblock{Parameters}{EndParameters}
\algnotext{EndParameters}
\newcommand{\Desc}[2]{\State \makebox[2em][l]{#1}#2}

\acmJournal{TCPS}
\acmVolume{0}
\acmNumber{0}
\acmArticle{0}
\acmYear{2019}
\acmMonth{0}
\copyrightyear{2019}

\setcopyright{acmlicensed}

\acmDOI{0000001.0000001}

\received{\today}

\renewcommand{\check}{\tilde}
\newcommand{\revise}[1]{{\color{black}#1}}

\begin{document}
\title[Implementing Homomorphic Encryption Based Secure Feedback Control]{Implementing Homomorphic Encryption Based Secure Feedback Control}

\author{Julian Tran}
\affiliation{%
  \institution{The University of Melbourne}
  \streetaddress{Grattan Street}
  \city{Parkville}
  \state{VIC}
  \postcode{3010}
  \country{Australia}}
\email{julian.tran@unimelb.edu.au}
\author{Farhad Farokhi}
\affiliation{%
  \institution{The University of Melbourne}
  \streetaddress{Grattan Street}
  \city{Parkville}
  \state{VIC}
  \postcode{3010}
  \country{Australia}}
\affiliation{%
  \institution{CSIRO's Data61}
  \city{Docklands}
  \state{VIC}
  \postcode{3008}
  \country{Australia}}
\email{ffarokhi@unimelb.edu.au}
\author{Michael Cantoni}
\affiliation{%
  \institution{The University of Melbourne}
  \streetaddress{Grattan Street}
  \city{Parkville}
  \state{VIC}
  \postcode{3010}
  \country{Australia}}
\email{cantoni@unimelb.edu.au}
\author{Iman Shames}
\affiliation{%
  \institution{The University of Melbourne}
  \streetaddress{Grattan Street}
  \city{Parkville}
  \state{VIC}
  \postcode{3010}
  \country{Australia}}
\email{ishames@unimelb.edu.au}

\begin{abstract}
This paper is about an encryption based approach to the secure implementation of feedback controllers
for physical systems. Specifically, Paillier's homomorphic encryption is used to digitally implement a class of linear dynamic controllers, which includes the commonplace static gain and PID type feedback control laws as special cases. The developed implementation is amenable to Field Programmable Gate Array (FPGA) realization. Experimental results, including timing analysis and resource usage characteristics for different encryption key lengths,
are presented for the realization of an inverted pendulum controller; as this is an unstable plant, the
control is necessarily fast.
\end{abstract}

%
%


%
%


\maketitle

\renewcommand{\shortauthors}{J. Tran et al.}

\section{Introduction}
\subsection{Motivation}
Advances in communication, control, and computer engineering have enabled the design and implementation of large-scale systems, such as smart infrastructure, with remote monitoring and control, which is often desired due to the geographical spread of the system and requirements for flexibility of design (to accommodate future expansions). These positive features however come at the cost of security threats and privacy invasions~\cite{teixeira2015secure,yang2015cost,qi2017demand, dong2018quantifying}.

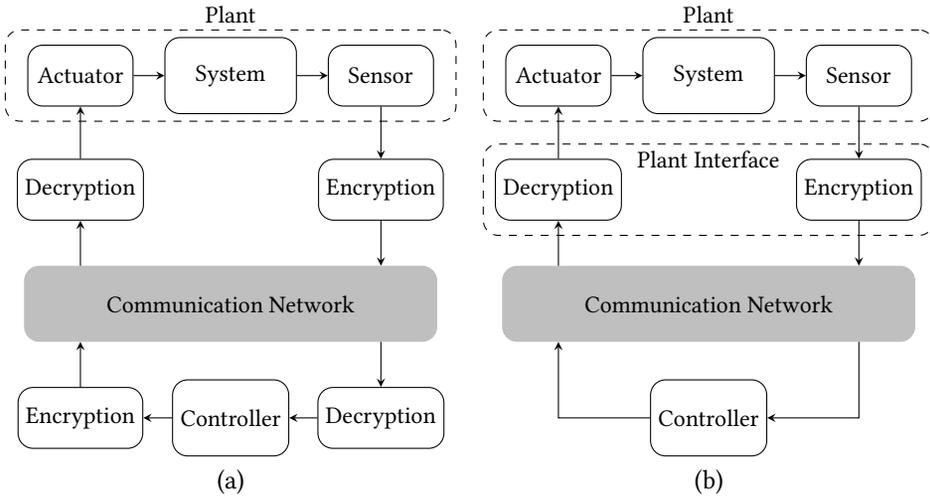
\begin{figure*}[t]
\centering
\begin{tabular}{cc}
\small
\begin{tikzpicture}[>=stealth]
\node[draw,rectangle,dashed,minimum width=6.cm,minimum height=1.2cm,rounded corners=0.2cm] (1) at (-0,-0) {};
\node[] at (-0.,+0.8) {Plant};
\node[draw,rectangle,minimum width=1.75cm,minimum height=1cm,rounded corners=0.2cm] (1) at (+0,-0) {System};
\node[draw,rectangle,minimum width=1.4cm,minimum height=.8cm,rounded corners=0.2cm] (2) at (+2.0,-0) {Sensor};
\node[draw,rectangle,minimum width=1cm,minimum height=.8cm,rounded corners=0.2cm] (3) at (+2.0,-1.5) {Encryption};
\node[draw,rectangle,minimum width=1.4cm,minimum height=.8cm,rounded corners=0.2cm] (4) at (-2.0,-0) {Actuator};
\node[draw,rectangle,minimum width=1.25cm,minimum height=1cm,rounded corners=0.2cm] (7) at (+0,-4.5) {Controller};
\node[draw,rectangle,minimum width=1cm,minimum height=.8cm,rounded corners=0.2cm] (10) at (-2.0,-1.5) {Decryption};
\node[rectangle,fill=gray!50,minimum width=5.5cm,minimum height=1cm,rounded corners=0.2cm] (N) at (0,-3) {Communication Network};
\node[draw,rectangle,minimum width=1cm,minimum height=.8cm,rounded corners=0.2cm] (5) at (-2.0,-4.5) {Encryption};
\node[draw,rectangle,minimum width=1cm,minimum height=.8cm,rounded corners=0.2cm] (6) at (+2.0,-4.5) {Decryption};
\draw[->] (1)  -- (2) ;
\draw[->] (2)  -- (3) ;
\draw[->] (3)  -- (+2.0,-2.5) ;
\draw[->] (+2.0,-3.5)  -- (6) ;
\draw[->] (6) -- (7) ;
\draw[->] (7)  -- (5) ;
\draw[->] (5)  -- (-2.0,-3.5) ;
\draw[->] (-2.0,-2.5)  -- (10);
\draw[->] (10) -- (4) ;
\draw[->] (4)  -- (1) ;
\end{tikzpicture}
&
\small
\begin{tikzpicture}[>=stealth]
\node[draw,rectangle,dashed,minimum width=6.cm,minimum height=1.2cm,rounded corners=0.2cm] (1) at (+0,-1.5) {};
\node[] at (0,-1.1) {Plant Interface};
\node[draw,rectangle,dashed,minimum width=6.cm,minimum height=1.2cm,rounded corners=0.2cm] (1) at (-0,-0) {};
\node[] at (-0.,+0.8) {Plant};
\node[draw,rectangle,minimum width=1.75cm,minimum height=1cm,rounded corners=0.2cm] (1) at (+0,-0) {System};
\node[draw,rectangle,minimum width=1.4cm,minimum height=.8cm,rounded corners=0.2cm] (2) at (+2.0,-0) {Sensor};
\node[draw,rectangle,minimum width=1cm,minimum height=.8cm,rounded corners=0.2cm] (3) at (+2.0,-1.5) {Encryption};
\node[draw,rectangle,minimum width=1.4cm,minimum height=.8cm,rounded corners=0.2cm] (4) at (-2.0,-0) {Actuator};
\node[draw,rectangle,minimum width=1.25cm,minimum height=1cm,rounded corners=0.2cm] (7) at (+0,-4.5) {Controller};
\node[draw,rectangle,minimum width=1cm,minimum height=.8cm,rounded corners=0.2cm] (10) at (-2.0,-1.5) {Decryption};
\node[rectangle,fill=gray!50,minimum width=5.5cm,minimum height=1cm,rounded corners=0.2cm] (N) at (0,-3) {Communication Network};
\draw[->] (1)  -- (2) ;
\draw[->] (2)  -- (3) ;
\draw[->] (3)  -- (+2.0,-2.5) ;
\draw[->] (+2.0,-3.5)  -- (+2.0,-4.5) -- (7) ;
\draw[->] (7)  -- (-2.0,-4.5) -- (-2.0,-3.5) ;
\draw[->] (-2.0,-2.5)  -- (10);
\draw[->] (10) -- (4) ;
\draw[->] (4)  -- (1) ;
\end{tikzpicture}
\\
(a) & (b)
\end{tabular}
\vspace*{-.15in}
\caption{\label{fig:schematic_diagram_ncs}The schematic diagram of a networked control system with (a) normal encryption and (b) semi-homomorphic encryption-decryption units.}
\vspace*{-.15in}
\end{figure*}

Security threats can be decomposed into multiple categories based on resources available to adversaries~\cite{shamesteixeira2015secure}. A basic security attack that requires relatively few resources is eavesdropping in which an adversary monitors communication links to extract valuable information about the underlying system. Eavesdropping is often a starting point for more sophisticated attacks~\cite{wang2013cyber}. These attacks have resulted in the use of encryption~\cite{wan2018physical, Patel2009ICS15387881538820}. Figure~\ref{fig:schematic_diagram_ncs}~(a) illustrates the schematic diagram of a typical secure cyber-physical system with encryption. The actuator, system, and sensor (sometimes together referred to as the plant) form the physical system that must be remotely monitored and controlled. The physical system can be the electricity grid, transportation network, or a building, for example. Note that, although a single node is used in Figure~\ref{fig:schematic_diagram_ncs}~(b) to denote the sensor, in general it can comprise a collection of spatially distributed sensors. That is, the sensors can be spread geographically within the underlying physical system to measure appropriate states in different locations, e.g., voltages and frequencies at various locations in an electricity grid. The same also goes for the actuator. The addition of the encryption and decryption units in Figure~\ref{fig:schematic_diagram_ncs}~(a) protects the overall system against eavesdroppers on the communication network; however, it does not provide any protection if the eavesdropper infiltrates the controller or if the controller itself is the eavesdropper (in industrial espionage). This is because sensitive information is decrypted prior to entering the controller and is thus readily available there. This motivates the use of a system, depicted in Figure~\ref{fig:schematic_diagram_ncs}~(b), with homomorphic encryption enabling controller computations to be performed on encrypted numbers.

In practice, the (physical) system in Figure~\ref{fig:schematic_diagram_ncs}~(b) is a continuous-time dynamical system. To control the system, the sensors sample the outputs of the system at regular intervals and transmit these measurements to the controller through communication networks (e.g., WiFi or Bluetooth for short ranges or the Internet for longer ranges). The controller computes the necessary commands based on the received measurements and forwards the commands to the actuators for implementation. The actuators then apply and hold the received control signal for a fixed duration. This methodology for digital control of physical systems is often, unsurprisingly, referred to as sample and hold~\cite{franklin1998digital}. Before each new sample can be processed by the controller, it must process the previous one, compute the control inputs, and transmit the control inputs to the actuators. Therefore, the sampling rate of the sensors cannot be faster than the inverse of the worst-case delay/latency caused by the required computations and communications. On the other hand, in order to guarantee stability and performance of the overall closed-loop system, we must ensure that the sampling occurs regularly and faster than a certain level (related to how fast the controlled systems dynamics need to be)~\cite{franklin1998digital}.

In~\cite{farokhi2017secure}, a general purpose microprocessor based system, specifically a Raspberry Pi, was used to control a differential-wheeled robot in real-time using an encrypted controller. Controlling such a robot is not a complicated task as the underlying system is stable and, if the control signal is not updated with regular timing, the system would not violate safety constraints so long as it is restricted to move very slowly. Further, slowing down the sampling rate in this robot only degrades the performance by making it slower, not resulting in undesirable behaviours. In safety critical applications, however, the timing of the control loop is crucial; if we cannot ensure that the controller is able to provide the correct actuation signal within the sampling time of the system, then safe operation of the system cannot be guaranteed. Having tight control on the timing is unfortunately not always possible on general purpose microprocessor based systems with operating systems because computations and their timings are subject to the operating system scheduling. Even without an operating system, the time sequential nature of software implementations for execution on a general purpose processor can
be limiting from the perspective of achievable sampling rate. This motivates the design of a custom digital engine, amenable to realization on Field-Programmable Gate Arrays (FPGAs), for performing the necessary computations. This is the focus of the developments presented below.
\vspace*{-.1in}
\subsection{Contributions}
In this paper, we use homomorphic encryption, specifically the Paillier encryption scheme~\cite{Paillier}, to implement linear control laws. This includes many popular control laws, such as static gain~\cite{green2012linear}, proportional-integral-derivative (PID) control~\cite{astrom2010feedback} and linear quadratic regulators (LQR)~\cite{green2012linear}. \revise{Linear control laws, such as PID controllers, have been heavily used within the industry for regulating nonlinear physical systems and are of practical relevance~\cite{aastrom2001future}.} Although the paper presents the digital system implementation within the context of Paillier encryption, the underlying methodology is applicable, in principle, to other homomorphic encryption methods that rely on the exponentiation of large integer numbers, such as RSA and ElGamal encryption~\cite{rsa,Elgamal}. After the quantization and transformation of the controllers for implementation on ciphertexts, modular multipliers and exponentiators are implemented using Montgomery multiplication~\cite{montgomery1985modular,koc}. These modules can be used in parallel for encryption, controller computations, and decryption. We analyze the timing of an FPGA realization of such an implantation of a feedback controller, and present experimental results for the control of an unstable system, namely, an inverted pendulum. 
\vspace*{-.1in}
\subsection{Related Studies}
The study of homomorphic encryption, a form of encryption that enables computations to be carried out on the encrypted data, dates back to the pioneering result of~\cite{rivest1978data} after observing semi-homomorphic properties in RSA~\cite{rsa}. \revise{Semi-homomorphic encryption only allows for a smaller number of operations to be performed on the encrypted data in contrast with fully homomorphic encryption.} For example, in the case of RSA and ElGamal encryption~\cite{Elgamal} multiplication of plaintext data corresponds to multiplication of encrypted data, and in the case of Paillier encryption~\cite{Paillier} summation of plaintext data corresponds to multiplication of encrypted data. The Gentry encryption scheme~\cite{gentry2009fully} is the first fully-homomorphic encryption scheme that allows both multiplication and summation of plain data through appropriate arithmetic operations on encrypted data. Subsequently, other fully homomorphic encryption methods have been proposed, e.g.,~\cite{van2010fully,brakerski2014leveled}. The computational burden of fully-homomorphic encryption methods is often much greater than that of semi-homomorphic encryption methods.

Homomorphic encryption has been used previously for third-party cloud-computing services~\cite{yu2018privacy,lopez2012fly,gentry2010computing, aono2016scalable,li2010secure,farokhi2017private}. More recent studies~\cite{shoukry2016privacy,farokhi2017secure,kogiso2015cyber, kim2016encrypting,kogiso2018upper} have considered challenges associated with the use of homomorphic encryption in closed-loop control of physical systems, such as maintaining stability and performance, albeit without considering timing concerns (by not getting into the computational time of encryption, computation, and decryption and assuming all underlying computations are instantaneous). None of these studies consider dynamic control laws; they are all restricted to static control laws without any form of memory. This is because, in dynamical control laws with an encrypted memory, the number of bits required for representing the state of the controller can grow linearly with the number of iterations. This renders the memory \revise{of such control laws} useless after a certain number iterations due to an overflow or an underflow\footnote{Underflow refers to the case where number of fractional bits required for representing a number becomes larger than the allowed number of fractional bits in a fixed-point number basis.}. We borrow theoretical results from~\cite{1100479,moore1973fixed,nevsic2008stability,murguia2018secure} to propose a finite-memory implementation of dynamic controllers over ciphertexts.

An alternative to homomorphic encryption is secure multi-party computation based on secret sharing or other forms of encryption (possibly non-homomorphic encryption methodologies). A well-known method for secure multi-party computation is the Yao protocol, which was originally developed for secure two-party computations~\cite{yao1982protocols}. The protocol provides a method for evaluating a Boolean function without any party being able to observe the bits that flow through the circuit during the evaluation. This has been proved to be secure~\cite{lindell2009proof} and efficiently implementable for Boolean functions~\cite{lindell2007efficient}. However, when dealing with more general mappings, i.e., non-Boolean functions, the efficiency of the protocol is limited as the problem of finding the most efficient Boolean representation of a function, in terms of the efficiency of implementing the Yao protocol~\cite{kolesnikov2008improved}, is not  trivial~\cite{kreuter2012billion}. Another approach is to utilize secret sharing in which a secret is divided into multiple shares and each party receives one share, which appears random to the receiving party. Then, appropriate computations on the secret shares can be performed to evaluate the outcome~\cite{chaum1988multiparty,kamm2015secure}. Application of secret sharing to general problems is difficult and the digital design becomes problem specific to the application.

Finally, note that the Paillier encryption scheme has been recently implemented on FPGAs in~\cite{san2016efficient}; however, that paper considered the problem of privacy-preserving data mining, which has different requirements in comparison to real-time encrypted control. This difference in requirements resulted in the consideration of a different implementation architecture in this paper. In particular, the binomial expansion for the specific choice of the exponential base is exploited to achieve fast encryption in this paper. Further, there are differences between the operations required for data mining and controller computation.
\vspace{-.1in}
\subsection{Paper Outline}
The rest of the paper is organized as follows. In Section~\ref{sec:secure},  the building blocks of the networked control systems in Figure~\ref{fig:schematic_diagram_ncs}~(b) are presented and we describe the implementation of the control laws over ciphertexts. In Section~\ref{sec:fpga}, the digital design for FPGA realization is described. We present the experimental results for the control of an inverted pendulum in Section~\ref{sec:experiments}. Finally, we conclude the paper and present avenues for future research in Section~\ref{sec:discussions}.
\vspace*{-.05in}
\section{Secure Feedback Control} \label{sec:secure}
In this section, we discuss encryption, decryption, and controller blocks of the networked control systems in  Figure~\ref{fig:schematic_diagram_ncs}~(b). 
\vspace*{-.1in}
\subsection{Feedback Controller}
In this paper, we consider dynamic controllers of the following form:\vspace*{-.05in}
\begin{subequations}\label{eq:dynamiccontroller}
\begin{align}
    \mathcal{C} :\quad
        x[k+1] &=
        \begin{cases}
            A x[k] + B (s[k] - y[k]), & k+1 \mod T > 0,\\
            0, & k+1 \mod T = 0,
        \end{cases}
        \\
        u[k] &= C x[k],
\end{align}
\end{subequations}
where $x[k] \in \mathbb{R}^{n_x}$ is the controller state, $u[k] \in \mathbb{R}^{n_u}$  is the vector of control inputs to the physical system, $y[k] \in \mathbb{R}^{n_y}$ is the vector of plant outputs, and $T$ is the number of time steps between controller state resets. Conditions for selecting $T$ with stability and performance guarantees are presented in~\cite{murguia2018secure}. The class of controllers in~\eqref{eq:dynamiccontroller} covers static, reset integral, reset lead and lag controller. For instance, in the case of static controllers, $A=0$, $B=I$, and $C$ is the static gain of the controller. Note that there is a delay of one sampling time between measurement and actuation, modelling computation and communication time associated with the networked controller. For static controllers, since the controller's state is not accumulative and only acts as a delay, we can set $T=\infty$ without concerns about state overflow or underflow. For reset proportional-integral (PI) controllers, $A=\diag(1,0)$ with $\diag(a)$ denoting a diagonal matrix whose main diagonal is equal to $a$, $B=[\Delta t \; 1 ]^\top$ with $\Delta t>0$ denoting the sampling time of the control system, and $C=[K_I \; K_p]$ with $K_I$ and $K_p$ denoting, respectively, the integral and proportional gains. Note that PI control laws have been heavily used within the industry for regulating/controlling nonlinear physical systems~\cite{aastrom2001future} and, therefore, the choice of linear dynamic controllers is of practical relevance. In this paper, we consider resetting dynamic control laws because implementing encrypted controllers over an infinite horizon is impossible due to memory issues (through repeated multiplication of fixed point numbers in the plaintext domain, the numbers of the bits required for representing the fractional and integer parts of plaintext numbers continuously grow, and there is no simple way to truncate with small error when working in the encrypted domain). Resetting controllers have been previously studied in~\cite{1100479,moore1973fixed,nevsic2008stability, murguia2018secure}.

\subsection{Homomorphic Encryption} \label{subsec:homomorphic}
A public key encryption scheme can be described by the tuple $(\mathbb{P}, \mathbb{C}, \mathbb{K}, \mathcal{E}, \mathcal{D})$, where $\mathbb{P}$ is the set of plaintexts, $\mathbb{C}$ is the set of ciphertexts, $\mathbb{K}$ is the set of keys, $\mathcal{E}$ is the encryption algorithm, and $\mathcal{D}$ is the decryption algorithm. As such encryption schemes are asymmetric, each key $\kappa = (\kappa _p, \kappa _s) \in \mathbb{K}$ is composed of a public key $\kappa _p$ (which is shared with everyone and is used to encrypt plaintexts), and a private key $\kappa _s$ (which is kept secret and is used to decrypt ciphertexts). The algorithms $\mathcal{E}$ and $\mathcal{D}$ are publicly known, and use the keys as parameters, which are generated for each new use-case. It is required that $\mathcal{D}(\mathcal{E}(x, \kappa _p), \kappa _p, \kappa _s) = x$.

\begin{definition}[Homomorphism in Cryptography] A public key encryption scheme $(\mathbb{P}, \mathbb{C}, \mathbb{K}, \mathcal{E}, \mathcal{D})$ is homomorphic if there exist operators $\circ$ and $\diamond$ such that $(\mathbb{P}, \circ)$ and $(\mathbb{C}, \diamond)$ are algebraic groups and $\mathcal{E}(x_1,\kappa_p)\diamond\mathcal{E}(x_2,\kappa_p) = \mathcal{E}(x_1 \circ x_2)$.
\end{definition}

Typically, the sets $\mathbb{P}$ and $\mathbb{C}$ are finite rings of integers $\mathbb{Z}_{n_P}$ and $\mathbb{Z}_{n_C}$ respectively. Then, the modular addition operation ($x_1 \circ x_2 = (x_1 + x_2) \mod n_P$) and the modular multiplication operation ($x_1 \circ x_2 = x_1x_2 \mod n_P$) both form groups with $\mathbb{P}$. If there exists an operator $\diamond$ that satisfies the definition of a homomorphic encryption scheme when $\circ$ is defined as modular addition, we call the encryption scheme additively homomorphic. Likewise, if there exists an operation $\diamond$ that satisfies the definition of a homomorphic encryption scheme when $\circ$ is defined as modular multiplication, we call the encryption scheme multiplicatively homomorphic. If both these properties hold, the encryption scheme is called fully-homomorphic; if only one description applies, it is semi-homomorphic. Importantly, the properties of fully-homomorphic and semi-homomorphic encryption schemes allow additions and multiplications of plaintexts to be performed through the generation of a ciphertext from other ciphertexts, without any intermediate decryptions and encryptions.

Encryption schemes, such as Paillier~\cite{Paillier}, RSA~\cite{rsa}, and ElGamal~\cite{Elgamal}, are examples of semi-homomorphic encryption. The Paillier encryption scheme is additively homomorphic, while the RSA and ElGamal encryption schemes are multiplicatively homomorphic. These homomorphic encryption schemes have been used in the literature to ensure privacy and security when various computational tasks, such as computing set intersections, data mining, executing arbitrary programs, and controlling dynamical systems, are performed by untrusted parties; see, e.g.,~\cite{yu2018privacy,lopez2012fly,gentry2010computing, aono2016scalable, li2010secure,farokhi2017secure,murguia2018secure} and references there-in for examples. The above-mentioned homomorphic encryption schemes involve calculating modular exponentiations (i.e., $b^a \mod M$ for positive integers $a$, $b$, and $M$), which is a computationally expensive operation. The time required to perform encryption, decryption, and homomorphic operations on ciphertexts, depends largely on the speed with which modular exponentiation can be achieved. This can potentially limit the usability of homomorphic encryption schemes for real-time control of physical systems.

\begin{definition}[Indistinguishability under Chosen Plaintext]
Consider a scenario in which a poly-nomial-time-bounded adversary provides two plaintexts. One of these plaintexts is randomly chosen and encrypted. An encryption scheme is said to be indistinguishable under chosen plaintext attack, if the adversary has a negligible advantage\footnote{Negligible advantage means that the difference between the probability of guessing the correct plaintext and the probability of guessing the wrong plaintext goes to zero rapidly as the key length goes to infinity~\cite{katz2014introduction}.} over guessing which of the two plaintexts were encrypted, using any information apart from the private key.
\end{definition}

Indistinguishability under chosen plaintext is a desirable property because an adversary is unable to determine the decryption of a ciphertext, by trialling encryption of likely plaintexts. The RSA encryption scheme does not have this property unless modified to OAEP-RSA~\cite{101007BFb0053428}. The Paillier and ElGamal encryption schemes have this property, as they introduce a large random number during encryption, allowing a single plaintext to encrypt non-deterministically to many possible ciphertexts, which removes any significant advantage in trialling encryption of likely plaintexts~\cite{Paillier,Elgamal}.

In what follows, we use Paillier encryption scheme as it is additively homomorphic and satisfies indistinguishability under chosen plaintext attack. Note that the ideas of this paper can be readily used for other homomorphic encryption relying on modular exponentiation. Paillier encryption works as follows. First, two large prime numbers $p$ and $q$ are randomly chosen to generate keys. The public key is $\kappa_p = N = pq$ and the private key is $\kappa_s = (\lambda, \mu) = (\lcm(p-1,q-1), \lambda^{-1} \mod N)$ where $\lcm(a, b)$ denotes the least common multiple of integers $a$ and $b$. Note that $\lambda^{-1}\mod N$ is a unique integer $\mu$ in $\mathbb{Z}_{N}$ such that $\lambda \mu \mod N=1$. In the Paillier encryption scheme, the set of plaintexts and ciphertexts are, respectively, $\mathbb{P} = \mathbb{Z}_{N}$ and $\mathbb{C} = \mathbb{Z}_{N^2}$. Encrypting a plaintext $t$ is done by calculating $\mathcal{E}(t) = (N+1)^t r^N \mod N^2$, where $r \in \{x \in \mathbb{Z}_N \mid \gcd(x, N) = 1\}$ is randomly chosen. Note that, because of using $N+1$ as the exponentiation basis in the encryption algorithm, it can be rewritten as $\mathcal{E}(t) = (Nt+1) r^N \mod N^2$. This property follows from the use of binomial expansion because
$
(N+1)^t r^N \mod N^2
=(\sum_{i=0}^t {t\choose i}N^i) r^N \mod N^2
=(Nt+1) r^N \mod N^2+(N^2\sum_{i=2}^t {t\choose i}N^{i-2}) r^N \mod N^2
=(Nt+1) r^N \mod N^2.
$
Using this property makes our implementation of the encryption considerably faster than~\cite{san2016efficient}. Decryption of a ciphertext $c$ is done by calculating $\mathcal{D}(c) = L(c^\lambda \mod N^2) \mu \mod N$, where $L(u) = (u-1)/N$.

The additive homomorphic property follows from $\mathcal{D}(\mathcal{E}(t_1, \kappa_p) \mathcal{E}(t_2, \kappa_p), \kappa_p, \kappa_s) = t_1 + t_2 \mod N$. Further, we have $\mathcal{D}(\mathcal{E}(t_1, \kappa_p)^{t_2}, \kappa_p, \kappa_s) = t_1 t_2 \mod N$. Note that this is not a true multiplicative homomorphic property, as $t_2$ is not encrypted; the encrypted result is formed from one ciphertext and one plaintext, rather than two ciphertexts. In the remainder of this paper, we use $\oplus$ to denote the additive homomorphic operator on ciphertexts and $\otimes$ to denote the pseudo-multiplicative homomorphic operator, i.e., 
\begin{subequations} \label{eqn:define_oplus_otimes}
\begin{align}
    \revise{c_1\oplus c_2:=}&\revise{(c_1c_2)\mod N^2},\\
    \revise{t\otimes c:=}&\revise{c^t\mod N^2.}
\end{align}
\end{subequations}
\subsection{Secure Controller Implementation}
The computations required to implement the controller in~\eqref{eq:dynamiccontroller} are additions and multiplications. We restrict the controller input to fixed-point numbers and use the mapping from fixed point numbers to the integers from \cite{farokhi2017secure}. This allows the equivalent operations of addition and multiplication to be effectively applied to fixed point numbers and integers over the ciphertext. The effect of the quantization error can be made arbitrarily small by increasing the number of bits used to represent the underlying numbers (specifically the number of fractional bits), at the expense of increased computational cost~\cite{farokhi2017secure,murguia2018secure}, given bounds on the size of disturbances that can act on the system. Quantizing also introduces saturation, which can be quite problematic. However, the negative effects of saturation may also be manged by increasing the number of bits (specifically the number of integer bits) used to represent the underlying numbers~\cite{farokhi2017secure,murguia2018secure}.

To provide more detail about the quantization process and its effect on the control law, we introduce the set of fractional numbers
\begin{align*}
\mathbb{Q}(n,m):=\bigg\{b\in\mathbb{Q}\,|\,b=- b_n2^{n-m-1}+\sum_{i=1}^{n-1}2^{i-m-1}b_i, b_i\in\{0,1\}\,\forall i\in\{1,\dots,n\}\bigg\}.
\end{align*}
The quantization operator $\mathcal{Q}:\mathbb{R}\rightarrow\mathbb{Q}$ is defined as $\mathcal{Q}(z):=\argmin_{z'\in\mathbb{Q}(n,m)} |z-z'|$. With slight abuse of notation, we use $\mathcal{Q}(A)$ and $\mathcal{Q}(x)$ to denote the entry-wise quantization of any $A\in\mathbb{R}^{n\times m}$ and $x\in \mathbb{R}^n$, respectively. The quantized controller is then given by
\begin{subequations}
\begin{align}
    \bar{\mathcal{C}} :\quad 
           \bar{x}[k+1] &=
        \begin{cases}
            \bar{A} \bar{x}[k] + \bar{B} (\bar{s}[k] - \bar{y}[k]), & k+1 \mod T > 0,\\
            0, & k+1 \mod T = 0,
        \end{cases}
        \\
        \bar{u}[k] &= \bar{C} \bar{x}[k],
\label{eq:quantized_dynamic_controller}
\end{align}
\end{subequations}
where $\bar{A}_{ij}=\mathcal{Q}(A_{ij})$, $\bar{B}_{ij}=\mathcal{Q}(B_{ij})$, $\bar{C}_{ij}=\mathcal{Q}(C_{ij})$, $\bar{s}_i[k]=\mathcal{Q}(s_i[k])$, and $\bar{y}_i[k]=\mathcal{Q}(y_i[k])$. We use the bar, e.g., $\bar{x}$, to denote the quantized version of any variable, e.g., $x$. The map from fixed point numbers to the integers $\mathbb{Z}_{2^{n'}}$ is borrowed from \cite{farokhi2017secure} to define
\begin{subequations}
\begin{align}
\hat{s}_i[k]&=(2^m \bar{s}_i[k])\mod 2^{n'},\\
\hat{y}_i[k]&=(2^m \bar{y}_i[k])\mod 2^{n'},\\
\hat{A}_{ij}&=(2^m \bar{A}_{ij})\mod 2^{n'},\\
\hat{B}_{ij}[k]&=(2^{(k\mod T+1)m} \bar{B}_{ij})\mod 2^{n'},\\
\hat{C}_{ij}&=(2^m \bar{C}_{ij})\mod 2^{n'},\\
\hat{x}_i[k]&=(2^{(k\mod T+1)m} \bar{x}_i[k])\mod 2^{n'},\\
\hat{u}_i[k]&=(2^{(k\mod T+2)m} \bar{u}_i[k+1])\mod 2^{n'},
\end{align}
\end{subequations}
where $n' = (n_x+1)T+n_u+n(T+2)$ to prevent overflows. Here, for simplicity, we assume that all scalar components of vectors use the same $n$ and $m$, but these values can differ for various parts of the controller in general~\cite{farokhi2017secure}. Then the quantized controller can then be rewritten to operate on ciphertexts as
\begin{subequations}\label{eq:dynamichom}
\begin{align}
    \tilde{\mathcal{C}} :
        \tilde{x}_i[k+1] &=
        \begin{cases}
            \left[\oplus _{j=1}^{n_x} (\hat{A}_{ij} \otimes \tilde{x}_j[k])\right]
            \oplus
            \left[\oplus _{j=1}^{n_y} (\hat{B}_{ij}[k] \otimes (\tilde{s}_j[k] - \tilde{y}_j[k]))\right], & k+1 \mod T > 0,\\
            \mathcal{E}(0,\kappa_p), & k+1 \mod T = 0,
        \end{cases}\\
        \tilde{u}_i[k] &= 
        \oplus _{j=1}^{n_x} (\hat{C}_{ij} \otimes \tilde{x}_j[k]),
\end{align}
\end{subequations}
where \revise{$\oplus,\otimes$ are defined in~\eqref{eqn:define_oplus_otimes} and} the tilde is used to denote the encrypted integers; i.e., $\tilde{u}_i[k]=\mathcal{E}(\hat{u}_i[k], \kappa_p)$,  $\tilde{s}_j[k]=\mathcal{E}(\hat{s}_j[k], \kappa_p)$,  $\tilde{y}_j[k]=\mathcal{E}(\hat{y}_j[k], \kappa_p)$,  $\tilde{x}_j[k]=\mathcal{E}(\hat{x}_j[k], \kappa_p)$. Finally, the control signal at the actuator is computed by
\begin{subequations}
\begin{align}
\hat{u}_i[k]&=\mathcal{D}(\tilde{u}_i[k],\kappa_p,\kappa_s)\mod 2^{n'},\\
\bar{u}_i[k]&=2^{-(k\mod T+2)m}(\hat{u}_i[k]-2^{n'}\mathds{1}_{\hat{u}_i[k]\geq 2^{n'-1}}),
\end{align}
\end{subequations}
where $\mathds{1}_{p}$ is equal to one if statement $p$ holds and is equal to zero otherwise. 

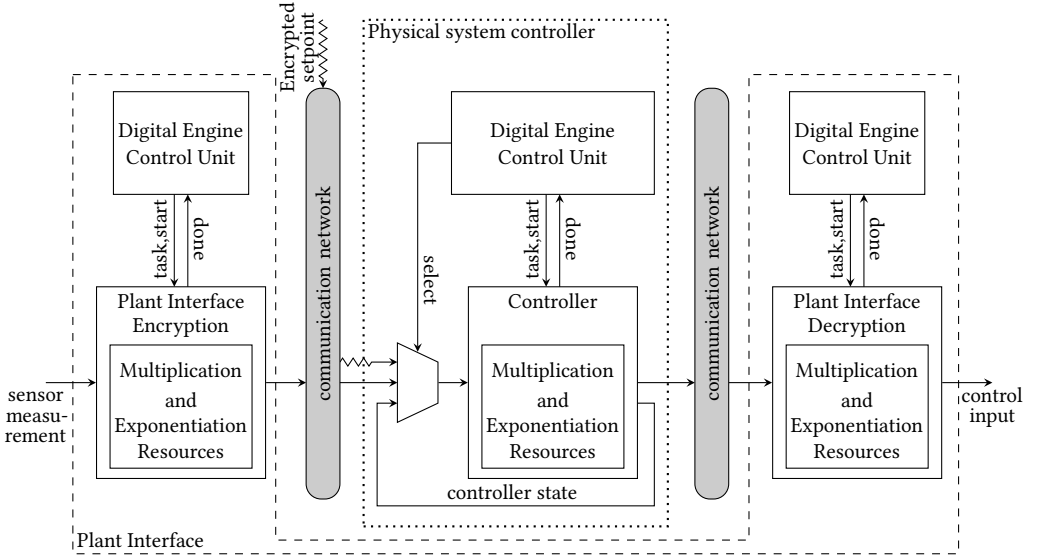
\begin{figure}
    \centering
    \newcommand\scaler{.9}
    \begin{tikzpicture}[>=stealth,x=1cm,y=1cm,scale=\scaler]
    \node[draw,rotate=90,trapezium,trapezium angle=-65,minimum height=\scaler*0.6cm] (M1) at (+2,0) {};    
    \node[draw,rectangle,minimum width=\scaler*2.5cm,minimum height=\scaler*2.8cm] (Enc) at (-1.5,0) {};
    \node[] at (-1.5,+1.20) {\footnotesize Plant Interface};
    \node[] at (-1.5,+0.80) {\footnotesize Encryption};
    \node[draw,rectangle,minimum width=\scaler*2.1cm,minimum height=\scaler*1.8cm] at (-1.5,-0.35) {};
    \node[] at (-1.5,+0.15) {\footnotesize Multiplication};
    \node[] at (-1.5,-0.25) {\footnotesize and};
    \node[] at (-1.5,-0.65) {\footnotesize Exponentiation};
    \node[] at (-1.5,-1.00) {\footnotesize Resources};
    \node[draw,rectangle,minimum width=\scaler*2.5cm,minimum height=\scaler*2.8cm] (Cr) at (+4,0) {};
    \node[] at (4,+1.20) {\footnotesize Controller};
    \node[draw,rectangle,minimum width=\scaler*2.1cm,minimum height=\scaler*1.8cm] at (+4,-0.35) {};
    \node[] at (4,+0.15) {\footnotesize Multiplication};
    \node[] at (4,-0.25) {\footnotesize and};
    \node[] at (4,-0.65) {\footnotesize Exponentiation};    
    \node[] at (4,-1.00) {\footnotesize Resources};
    \node[draw,rectangle,minimum width=\scaler*2.5cm,minimum height=\scaler*2.8cm] (Dec) at (+8.5,0) {};
    \node[] at (8.5,+1.20) {\footnotesize Plant Interface};
    \node[] at (8.5,+0.80) {\footnotesize Decryption};
    \node[draw,rectangle,minimum width=\scaler*2.1cm,minimum height=\scaler*1.8cm] at (+8.5,-0.35) {};
    \node[] at (8.5,+0.15) {\footnotesize Multiplication};
    \node[] at (8.5,-0.25) {\footnotesize and};
    \node[] at (8.5,-0.65) {\footnotesize Exponentiation};   
    \node[] at (8.5,-1.00) {\footnotesize Resources};
    \node[draw,rectangle,minimum width=\scaler*3.0cm,minimum height=\scaler*1.5cm] at (4,3.5) {
    \begin{minipage}{2.0cm}
    \centering
    \footnotesize 
    Digital Engine \\ Control Unit
    \end{minipage}
    };
    \node[draw,rectangle,minimum width=\scaler*2.cm,minimum height=\scaler*1.5cm] at (8.5,3.5) {};
    \node[] at (8.5,3.5) {
    \begin{minipage}{2cm}
    \centering
    \footnotesize
    Digital Engine \\ Control Unit
    \end{minipage}
    };    
    \node[draw,rectangle,minimum width=\scaler*2.0cm,minimum height=\scaler*1.5cm] at (-1.5,3.5) {};    
    \node[] at (-1.5,3.5) {
    \begin{minipage}{2.cm}
    \centering
    \footnotesize
    Digital Engine \\ Control Unit
    \end{minipage}
    };
    \draw[-,dotted,thick] (1.2,5.3) -- (5.7,5.3) -- (5.7,-2.1) -- (1.2,-2.1)  -- cycle;
    \draw[-,dashed] (-3.1,4.5) -- (-0.1,4.5) -- (-0.1,-2.3) -- (6.9,-2.3) -- (6.9,4.5) -- (10,4.5) -- (10,-2.5) -- (-3.1,-2.5) -- cycle;
    \draw[->] (0.75,0.0) -- (1.7,0.0);
    \draw[->] (M1) -- (Cr);
    \draw[->] (5.25,-0.3) -- (5.50,-0.3) -- (5.50,-1.8) -- (1.40,-1.8) -- (1.40,-0.3) -- (1.7,-0.3);
    \draw[decoration={aspect=0.3, segment length=4pt, amplitude=1.5pt,zigzag,post length=8pt},decorate,->] (.85,0.3) -- (1.7,0.3);
    \draw[decoration={aspect=-0.3, segment length=4pt, amplitude=-1.5pt,zigzag,post length=3pt},decorate,->] (.6,5.3) -- (.6,4.3);
    \draw[->] (Cr) -- (6.1,0);
    \draw[->] (6.6,0) -- (Dec);
    \draw[->] (Enc) -- (.35,0);
    \draw[->] (-3.5,0) -- (Enc);
    \draw[->] (Dec) -- (10.5,0);
    \draw[<-] (-1.4,2.75) -- (-1.4,1.4);
    \node[rotate=+90] at (-1.8,2.1) {\footnotesize task,start};
    \draw[->] (-1.6,2.75) -- (-1.6,1.4);
    \node[rotate=-90] at (-1.2,2.1) {\footnotesize done};
    \draw[->] (3.9,2.75) -- (3.9,1.4);
    \node[rotate=+90] at (3.7,2.1) {\footnotesize task,start};
    \draw[<-] (4.1,2.75) -- (4.1,1.4);
    \node[rotate=-90] at (4.3,2.1) {\footnotesize done};
    \draw[->] (8.4,2.75) -- (8.4,1.4);
    \node[rotate=+90] at (8.2,2.1) {\footnotesize task,start};
    \draw[<-] (8.6,2.75) -- (8.6,1.4);
    \node[rotate=-90] at (8.8,2.1) {\footnotesize done};
    \draw[->] (2.5,3.5) -- (2.0,3.5) -- (2.0,0.45);
    \node[rotate=-90] at (2.2,1.5) {\footnotesize select};
    \node[] at (3.4,-1.6) {\footnotesize controller state};
    \node[] at (-2.1,-2.3) {\footnotesize Plant Interface};
    \node[] at (2.95,5.1) {\footnotesize Physical system controller};
    \node[] at (10.50,-0.35) {
    \begin{minipage}{1.5cm}
    \centering
    \footnotesize 
    control \\[-.3em] input
    \end{minipage}
    };
    \node[] at (-3.65,-0.5) {
    \begin{minipage}{1.5cm}
    \centering
    \footnotesize 
    sensor 
    \\[-.3em] measu-
    \\[-.3em] rement
    \end{minipage}
    };
    \node[rotate=90] at (.4,+4.9) {\footnotesize setpoint};
    \node[rotate=90] at (.1,+4.9) {\footnotesize Encrypted};
    \node[draw,rotate=90,rounded corners=2mm,minimum height=\scaler*.5cm,minimum width=\scaler*6cm,fill=black!20] at (0.6,1.3) {\footnotesize communication network};
    \node[draw,rotate=90,rounded corners=2mm,minimum height=\scaler*.5cm,minimum width=\scaler*6cm,fill=black!20] at (6.35,1.3) {\footnotesize communication network};
    \end{tikzpicture}
    \caption{Schematic diagram of the custom digital system for encrypted control.}
    \label{fig:my_label}
\end{figure}

\section{Digital Design} \label{sec:fpga}
Timing is an important issue when implementing controllers in real-time. While the maximum computation to be performed by the controller is effectively the same in every iteration, implementations on a general purpose microprocessor based system are subject to variable timing performance dependent on operating system scheduling.  Even without an operating system, the time sequential nature of software implementations for execution on a general purpose processor can be limiting from the perspective of achievable sampling rate. Such implementations are therefore not acceptable for systems with strict deadlines. This motivates the development of a custom digital engines for performing the computations. Hardware implementation of homomorphic encryption based secure feedback control can result in faster sampling rates than software implementations, thereby broadening the applicability of encryption based methods for securing feedback control systems.
The speedup of a digital design in hardware over a software design can be from many aspects. Hardware designs are able to take advantage of full parallelism, while software designs typically run sequentially on a few parallel threads, and are thus limited in their parallelism. Hardware designs can also introduce pipelining into data paths, where the computation is divided into a pipeline of sequential stages, with stages all running at the same time, and each stage passing its result to the next stage~\cite{ward1990computation}. This can be used to increase achievable data throughput compared to sequential software designs, as new data can be passed through the first stage of the pipeline while there is still data to be processed in the subsequent stages.

\begin{algorithm}[t]

\begin{algorithmic}[1]
    \Parameters
        \Desc{$N$}{Paillier public key}
        \Desc{$R$}{Montgomery radix}
    \EndParameters
    \Inputs
        \Desc{$B$}{Integer base in Montgomery form $B = bR \mod N^2$}
        \Desc{$E$}{Integer exponent with $l$ bits}
    \EndInputs
    \Outputs
        \Desc{$P$}{Power in Montgomery form $P = b^ER \mod N^2$}
    \EndOutputs
    \Function{MontExp}{$B, E$}
        \State $P\leftarrow R \mod N^2$
        \For{$i = 1, ..., l$}
            \If{$E \mod 2 = 1$}
                \State $P \leftarrow \textproc{MontMult}[M=N^2](P, B)$
            \EndIf
            \State $E \leftarrow \floor{E / 2}$
            \State $B \leftarrow \textproc{MontMult}[M=N^2](B, B)$
        \EndFor
        \State \Return $P$
    \EndFunction
\end{algorithmic}

\caption{Right-to-left method for modular exponentiation using Montgomery multiplication in modulus $M=N^2$}
\label{alg:rightleftmod}
\end{algorithm}

Figure~\ref{fig:my_label} illustrates the schematic diagram of the custom digital system for encrypted control discussed in this section. There are three major parts: encryption and decryption units, in the plant interface, and the physical system controller unit, accessed
over a network. Each of these units includes a digital engine controller, which orchestrates data flow through the components of these systems, according to a corresponding algorithmic state machine. The activity of each major part is triggered by external events. Encryption is periodically triggered by the generation of samples of the plant output. Physical system controller and decryption unit activity is triggered by the arrival of data over the network.

In this section, we describe plant interface (encryption and decryption) and physical system controller blocks in Figure~\ref{fig:my_label}. Modular multiplication and modular exponentiation are important recurring elements in all of these blocks. Therefore, we start by describing these elemental building blocks in Subsection~\ref{subsection:exponentiation}. We then describe the controller in Subsection~\ref{subsec:controller} and the plant interface in Subsection~\ref{subsec:plantinterface}.

\subsection{Modular Multiplication and Exponentiation} \label{subsection:exponentiation}

In many homomorphic encryption schemes, including Paillier encryption, efficient implementation of modular exponentiation is essential for fast encryption, decryption, and homomorphic operations; see Subsection~\ref{subsec:homomorphic}. Within the context of secure feedback control implementation, the time it takes to perform encryption,
decryption and homomorphic operations on cyphertexts, is a lower bound on the control loop sample
period, which when reduced, typically leads to improved performance for systems with fast dynamics
(e.g., an unstable inverted pendulum).
Note that, in principle, it is possible to decrease the time required for computations by decreasing the encryption key length; however, this would reduce the security of the system which is not desirable.

\begin{algorithm}[t]

\begin{algorithmic}[1]
    \Parameters
        \Desc{$M$}{Odd modulus}
        \Desc{$w$}{Number of 16 bit words such that $M < 2^{16w}$}
        \Desc{$M'$}{such that $MM' \mod 2^{16} = 2^{16} - 1$}
    \EndParameters
    \Inputs
        \Desc{$X$}{Input such that $X < 2M$}
        \Desc{$Y$}{Input such that $Y < 2M$}
    \EndInputs
    \Outputs
        \Desc{$T$}{Such that $T \mod M = XYR^{-1} \mod M, T < 2M$, where $R = 2^{16(w+1)}$, $RR^{-1} \mod M = 1$}
    \EndOutputs
    \Function{MontMult}{$X, Y$}
        \State $T = 0$
        \For{$i = 1, ..., w + 1$}
            \State $Z \leftarrow X(Y \mod 2^{16})$
            \State $Y \leftarrow \floor{Y / 2^{16}}$
            \State $m \leftarrow ((T \mod 2^{16}) + (Z \mod 2^{16}))M' \mod 2^{16}$
            \State $T \leftarrow (T + Z + mM) / 2^{16}$
        \EndFor
        \State \Return $T$
    \EndFunction
\end{algorithmic}

\caption{\cite{final-subtraction2} Modified Coarsely Integrated Operand Scanning (CIOS) method variant of Montgomery multiplication using 16 bits per word and without the  final conditional subtraction.}
\label{alg:cios2}
\end{algorithm}

\begin{figure}
    \centering
    \newcommand\scaler{.9}
    \begin{tikzpicture}[>=stealth,x=1cm,y=1cm,scale=\scaler]
    \node[draw,rectangle,minimum width=\scaler*2.6cm,minimum height=\scaler*1.7cm] at (0,0) {
    \begin{minipage}{2cm}
    \centering
    \footnotesize
    Modular \\
    multiplication
    \end{minipage}
    };
    \node[draw,rotate=90,trapezium,trapezium angle=-65,minimum height=\scaler*0.6cm] at (+2.3,1) {};    
    \draw[->] (1.3,0) -- (1.65,0) -- (1.65,.7) -- (2,.7);
    \draw[->] (-3.8,1.3) -- (2,1.3);
    \draw[->] (2.6,1) -- (2.9,1) -- (2.9,-1.2) -- (-2.2,-1.2) -- (-2.2,0) -- (-1.8,0) -- (-1.8,0.4) -- (-1.3,0.4);
    \draw[->] (-1.8,0) -- (-1.8,-0.4) -- (-1.3,-0.4);
    \begin{scope}[xshift=.6cm]
    \node[draw,rectangle,minimum width=\scaler*3.6cm,minimum height=\scaler*.7cm] at (0,-2.3) {\footnotesize memory: power};
    \node[draw,rectangle,minimum width=\scaler*2.6cm,minimum height=\scaler*1.7cm] at (.5,-3.8) {
    \begin{minipage}{2cm}
    \centering
    \footnotesize
    Modular \\
    multiplication
    \end{minipage}
    };    
    \node[draw,rotate=90,trapezium,trapezium angle=-65,minimum height=\scaler*0.6cm] at (+2.8,-2.75) {};    
    \draw[->] (1.8,-2.3) -- (2.05,-2.3) -- (2.05,-2.45) -- (2.5,-2.45);
    \draw[->] (1.8,-3.8) -- (2.05,-3.8) --  (2.05,-3.05) -- (2.5,-3.05);
    \draw[->] (3.1,-2.75) -- (3.4,-2.75) -- (3.4,-1.7) -- (-2.2,-1.7) -- (-2.2,-2.3) -- (-1.8,-2.3);
    \draw[->] (-2.2,-2.3) -- (-2.2,-3.5) -- (-0.8,-3.5);
    \draw[->] (-2.8,-1.2) -- (-2.8,-4.1) -- (-0.8,-4.1);
    \node[draw,rectangle,minimum width=\scaler*4.6cm,minimum height=\scaler*.7cm] at (0,-5.4) {\hspace{-.4in}\raisebox{3pt}{\footnotesize shift register: exponent}};
    \draw[->] (-2.2,-5.6) -- (+1.3,-5.6) ;
    \draw[- ] (2,-5.75) -- (2,-5.05); 
    \draw[- ] (1.7,-5.75) -- (1.7,-5.05); 
    \draw[- ] (1.4,-5.75) -- (1.4,-5.05); 
    \draw[->] (2.3,-5.3) -- (2.8,-5.3) -- (2.8,-3.2);
    \draw[->] (2.3,-5.5) -- (4.8,-5.5);
    \end{scope}
    \draw[dashed] (-3,1.8) -- (-3,-6) -- (4.5,-6) -- (4.5,1.8) -- cycle;
    \draw[->] (-3.8,-5.5) -- (-1.7,-5.5);
    \node[] at (-3.6,-5.3) {\footnotesize $E$};
    \node[] at (-3.6,+1.45) {\footnotesize $B$};
    \draw[- ] (2.65,-2.3) -- (2.65,-2.0) -- (3.9,-2.0);
    \draw[- ] (3.9,-2.0) arc (-180:0:0.1);
    \draw[->] (4.1,-2.0) -- (5.3,-2.0);
    \draw[->] (2.3,2.5) -- (2.3,1.45);
    \draw[- ] (2.3,1.67) -- (-2.5,1.67) -- (-2.5,1.4);
    \draw[- ] (-2.5,1.4) arc (+90:-90:0.1);
    \draw[->] (-2.5,1.2) -- (-2.5,-5.3) -- (-1.7,-5.3);
    \node[rotate=-90] at (2.5,2.2) {\footnotesize start};
    \node[] at (5.35,-1.75) {\footnotesize $B^E\mod N^2$};
    \node[] at (4.9,-5.3) {\footnotesize done};
    \end{tikzpicture}
    \caption{Block diagram of the Montgomery exponentiator using two modular Montgomery multipliers.}
    \label{fig:multiply_to_exponential}
\end{figure}
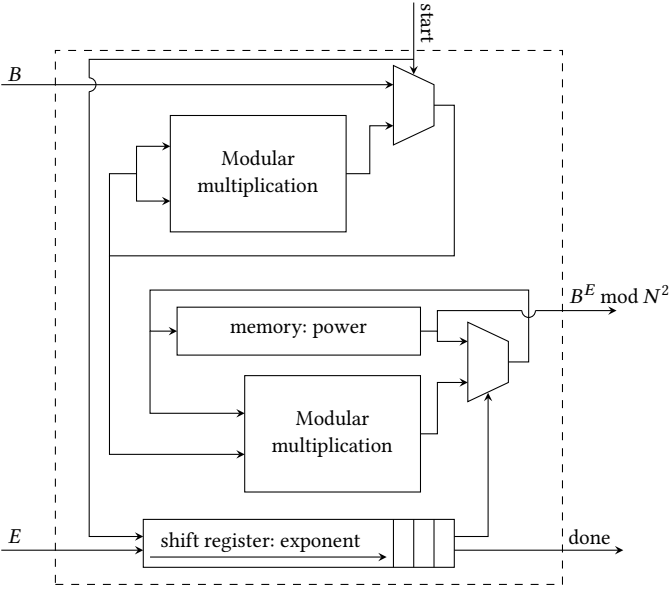

We utilize the right-to-left binary method for calculating modular exponentiation, which is summarized in Algorithm~\ref{alg:rightleftmod}. The algorithm is particularly useful for our application as it allows for the parallelization of the two modular multiplications in each iteration. This gives a speedup of up to two times, and results in a constant latency as the modular multiplication \revise{in line 13 in Algorithm~\ref{alg:rightleftmod}} is performed in parallel to the modular multiplication that must be always performed in each iteration \revise{in line 16 in Algorithm~\ref{alg:rightleftmod}}. The right-to-left binary method for exponentiation involves calculating many sequential modular multiplications. The algorithm best suited for this purpose is Montgomery multiplication~\cite{montgomery1985modular}. It removes the need to perform a trial division by the modulus which is an expensive operation in hardware, and instead only involves additions, multiplications, and right shifts; e.g., see Algorithm~\ref{alg:cios2}. However, for it to be useful for implementing modular multiplications, its operands must be converted to Montgomery form, and the result must be converted back from Montgomery form. These conversions can be done using additional Montgomery multiplications. The Montgomery form of an integer $a$ when using a modulus of $M$ is $(aR) \mod M$, where the Montgomery radix $R$ is typically a power of $2$, larger than $M$. 
In the right-to-left binary implementation of modular exponentiation, subsequently, referred to as Montgomery exponentiation, the conversions to and from the Montgomery form only occur before and after the exponentiation, as the intermediate (theoretical) conversions between the sequential multiplications within the exponentiation cancel out~\cite{montgomery1985modular}. The block diagram for a realization of the Montgomery exponentiator is illustrated in Figure~\ref{fig:multiply_to_exponential}. 

Many hardware designs for computing Montgomery multiplications exist. A design involving the Karatsuba multiplication algorithm can be used to evaluate very large multiplications~\cite{karatsubamont}. While this proved to be computationally effective in \cite{karatsubamont}, such a method may not be suitable for some applications due to prohibitive hardware resource required for evaluating Montgomery multiplications even with relatively small operands. Another method for implementing Montgomery multiplication involves using the Coarsely Integrated Operand Scanning (CIOS) variant \cite{koc} with a word size of a single bit. Implementations of this algorithm are described in \cite{radix2_1, radix2_2}. The bitwise approach greatly simplifies the architecture of the Montgomery multiplier, as it is only required to perform additions and right shifts. However, the bitwise design cannot make use of the multi-bit word embedded multipliers available on most modern FPGA devices.

A blockwise implementation of the CIOS method of Montgomery multiplication is ideal for the purposes of this paper as it is amenable to the use of embedded multipliers in FPGAs to perform smaller multiplications. Some implementations of this algorithm are discussed in \cite{thesis, highradix_1, highradix_2}. These implementations range from using a constant number of embedded multipliers to the case where the number of embedded multipliers scales linearly with the number of bits in the operands to perform large parallel multiplications. Therefore, based on the amount of the available hardware resources, an appropriate implementation of the blockwise CIOS-based Montgomery multiplier can be designed to ensure the resources are utilized effectively. 

In Algorithm~\ref{alg:cios2}, we borrow the modified CIOS method \cite{final-subtraction2} with a word size of 16 bits. The modified CIOS method removes the conditional final subtraction in typical Montgomery multiplication implementations to reduce hardware resource consumption.
Algorithm~\ref{alg:cios2} also differs from the conventional Montgomery multiplication in that it produces outputs that possibly have the modulus $M$ added to it, rather than an output in $\mathbb{Z}_M$. \revise{Such an output} is acceptable as long as an explicit conversion from this modified Montgomery form, through Montgomery multiplication by $1$, is used to produce the final result~\cite{final-subtraction2}.

\begin{figure}
    \centering
\begin{tikzpicture}[>=stealth,scale=.85]
        \newcommand{\scaler}{.85}
        \begin{scope}[xshift=-100mm,yshift=+0mm,
        yshift=90,every node/.append style={
        yslant=0.5,xslant=-1},yslant=0.5,xslant=-1]
            \draw[-,dashed,thick,fill=white] (0.8,1.6) -- (6,1.6) -- (6,-2.1) -- (0.8,-2.1)  -- cycle;
            \node[draw,rotate=90,trapezium,trapezium angle=-65,minimum height=\scaler*0.6cm] (M1) at (+2,0) {};    
            \node[draw,rectangle,minimum width=\scaler*2.5cm,minimum height=\scaler*2.8cm] (Cr) at (+4,0) {};
            \node[] at (4,+0.15) {\footnotesize Multiplication};
            \node[] at (4,-0.25) {\footnotesize and};
            \node[] at (4,-0.65) {\footnotesize Exponentiation}; 
            \draw[->] (0.25,0.0) -- (1.7,0.0);
            \draw[->] (M1) -- (Cr);
            \draw[->] (5.25,-0.3) -- (5.50,-0.3) -- (5.50,-1.8) -- (1.40,-1.8) -- (1.40,-0.3) -- (1.7,-0.3);
            \draw[decoration={aspect=0.3, segment length=4pt, amplitude=1.5pt,zigzag,post length=8pt},decorate,->] (.25,0.3) -- (1.7,0.3);
            \draw[->] (Cr) -- (6.7,0);
            \draw[->] (2.0,2.25) -- (2.0,0.45);
        \end{scope}
        \begin{scope}[xshift=-100mm,yshift=+5mm,
        yshift=90,every node/.append style={
        yslant=0.5,xslant=-1},yslant=0.5,xslant=-1]
            \draw[-,dashed,thick,fill=white] (0.8,1.6) -- (6,1.6) -- (6,-2.1) -- (0.8,-2.1)  -- cycle;
            \node[draw,rotate=90,trapezium,trapezium angle=-65,minimum height=\scaler*0.6cm] (M1) at (+2,0) {};    
            \node[draw,rectangle,minimum width=\scaler*2.5cm,minimum height=\scaler*2.8cm] (Cr) at (+4,0) {};
            \node[] at (4,+0.15) {\footnotesize Multiplication};
            \node[] at (4,-0.25) {\footnotesize and};
            \node[] at (4,-0.65) {\footnotesize Exponentiation}; 
            \draw[->] (0.25,0.0) -- (1.7,0.0);
            \draw[->] (M1) -- (Cr);
            \draw[->] (5.25,-0.3) -- (5.50,-0.3) -- (5.50,-1.8) -- (1.40,-1.8) -- (1.40,-0.3) -- (1.7,-0.3);
            \draw[decoration={aspect=0.3, segment length=4pt, amplitude=1.5pt,zigzag,post length=8pt},decorate,->] (.25,0.3) -- (1.7,0.3);
            \draw[->] (Cr) -- (6.7,0);
            \draw[->] (2.0,2.25) -- (2.0,0.45);
        \end{scope}     
        \begin{scope}[xshift=-100mm,yshift=+10mm,
        yshift=90,every node/.append style={
        yslant=0.5,xslant=-1},yslant=0.5,xslant=-1]
            \draw[-,dashed,thick,fill=white] (0.8,1.6) -- (6,1.6) -- (6,-2.1) -- (0.8,-2.1)  -- cycle;
            \node[draw,rotate=90,trapezium,trapezium angle=-65,minimum height=\scaler*0.6cm] (M1) at (+2,0) {};    
            \node[draw,rectangle,minimum width=\scaler*2.5cm,minimum height=\scaler*2.8cm] (Cr) at (+4,0) {};
            \node[] at (4,+0.15) {\footnotesize Multiplication};
            \node[] at (4,-0.25) {\footnotesize and};
            \node[] at (4,-0.65) {\footnotesize Exponentiation}; 
            \draw[->] (0.25,0.0) -- (1.7,0.0);
            \draw[->] (M1) -- (Cr);
            \draw[->] (5.25,-0.3) -- (5.50,-0.3) -- (5.50,-1.8) -- (1.40,-1.8) -- (1.40,-0.3) -- (1.7,-0.3);
            \draw[decoration={aspect=0.3, segment length=4pt, amplitude=1.5pt,zigzag,post length=8pt},decorate,->] (.25,0.3) -- (1.7,0.3);
            \draw[->] (Cr) -- (6.7,0);
            \draw[->] (2.0,2.25) -- (2.0,0.45);
        \end{scope}
        \begin{scope}[xshift=-100mm,yshift=+20mm,
        yshift=90,every node/.append style={
        yslant=0.5,xslant=-1},yslant=0.5,xslant=-1]
            \draw[-,dashed,thick,fill=white] (0.8,1.6) -- (6,1.6) -- (6,-2.1) -- (0.8,-2.1)  -- cycle;
            \node[draw,rotate=90,trapezium,trapezium angle=-65,minimum height=\scaler*0.6cm] (M1) at (+2,0) {};    
            \node[draw,rectangle,minimum width=\scaler*2.5cm,minimum height=\scaler*2.8cm] (Cr) at (+4,0) {};
            \node[] at (4,+0.15) {\footnotesize Multiplication};
            \node[] at (4,-0.25) {\footnotesize and};
            \node[] at (4,-0.65) {\footnotesize Exponentiation}; 
            \draw[->] (0.25,0.0) -- (1.7,0.0);
            \draw[->] (M1) -- (Cr);
            \draw[->] (5.25,-0.3) -- (5.50,-0.3) -- (5.50,-1.8) -- (1.40,-1.8) -- (1.40,-0.3) -- (1.7,-0.3);
            \draw[decoration={aspect=0.3, segment length=4pt, amplitude=1.5pt,zigzag,post length=8pt},decorate,->] (.25,0.3) -- (1.7,0.3);
            \draw[->] (Cr) -- (6.7,0);
            \draw[->] (2.0,2.25) -- (2.0,0.45);
        \end{scope}
        \begin{scope}[xshift=-67mm,yshift=+1mm]
        \node[] at (-4.2,5.9) {\huge $\vdots$};
        \draw[-,thick] (-4.5,6.5) -- (-4.7,6.5) -- (-4.7,3.9) -- (-4.5,3.9);
        \node[rotate=90] at (-5,5.3) {$n_y$ copies};
        \end{scope}
\end{tikzpicture}
    \caption{Block diagram of control computation using the Montgomery multiplication and exponentiation. These parallel blocks sit within the controller block in Figure~\ref{fig:my_label}. }
    \label{fig:encryption_diagram}
\end{figure}
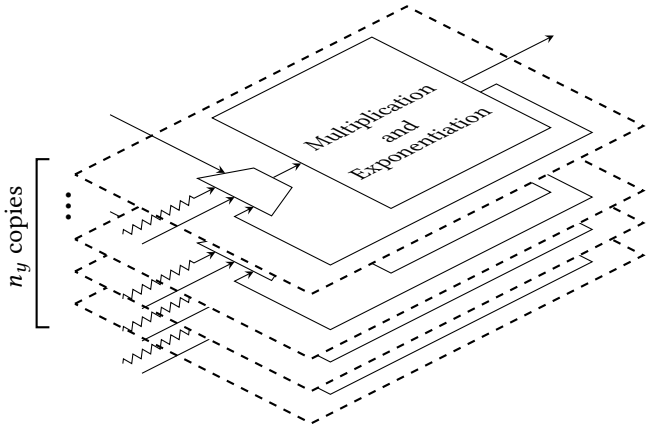

Across all Montgomery multipliers, we use the same value of the Montgomery radix $R = 2^{16(w+1)}$, where $w$ is the smallest integer such that $N^2 + 2 < 2^{16w}$; note that $N^2 + 2$ is the largest modulus used in the system. Throughout the encrypted control system, there are only three different values used as modulus, so these values can be coded into the Montgomery multipliers required, with an input allowing for the selection of the modulus. In the Paillier encryption scheme, all modular exponentiations have modulus $M = N^2$, where $N$ is the public key.

In what follows, using the custom digital implementations of the Montgomery multipliers and the Montgomery exponentiators as the underlying arithmetic blocks, we design plant interface and physical system controller modules for an encrypted control system secured with the Paillier encryption scheme. As shown in Figure ~\ref{fig:schematic_diagram_ncs}, the plant interface performs encryptions of system outputs and decryptions of control inputs, and the controller evaluates the control law securely over encrypted data. The ciphertexts transmitted between the plant interface and the controller are in the Montgomery form.

\begin{algorithm}[t]
\begin{algorithmic}[1]
    \Parameters
        \Desc{$N$}{Paillier public key}
        \Desc{$R$}{Montgomery radix}
        \Desc{$n'$}{Number of bits in mapping from fixed point numbers $\mathbb{Q}(n, m)$ to integers $\mathbb{Z}_{2^{n'}}$}
        \Desc{$\hat{C}$}{Controller matrix}
    \EndParameters
    \Inputs
        \Desc{$\check{x}$}{Encrypted controller state in Montgomery form $\check{x}_1, ..., \check{x}_{n_x}$}
    \EndInputs
    \Outputs
        \Desc{$\check{u}$}{Encrypted control inputs in Montgomery form $\check{u}_1, ..., \check{u}_{n_u}$}
    \EndOutputs
    
    \Function{GenerateControl}{$\check{y}, \check{s}, \check{x}$}
        \For{$i = 1, ..., n_u$} \Comment{Generate encrypted scalar products}
            \For{$j = 1, ..., n_x$}
                \State $\mathrm{var}_{ij} \leftarrow \textproc{MontExp}(\check{x}_j, \hat{C}_{ij})$
            \EndFor
        \EndFor
        \For{$i = 1, ..., n_u$} \Comment{Homomorphically sum up encrypted scalar products}
            \For{$j = 2, ..., n_x$}
                \State $\mathrm{var}_{i1} \leftarrow \textproc{MontMult}[M=N^2](\mathrm{var}_{i1}, \mathrm{var}_{ij})$
            \EndFor
            \State $\check{u}_i \leftarrow \mathrm{var}_{i1}$
        \EndFor
        \State \Return $\check{u}$
    \EndFunction
\end{algorithmic}
\caption{Computing the control input using the Montgomery multiplication and the Montgomery exponentiation.}
\label{alg:dynamiccontrol}
\end{algorithm}

Parallelization is possible within the building blocks of the Montgomery multiplier and the Montgomery exponentiator, and also in the designs of the plant interface and controller. Adding parallelization increases the resource consumption of the hardware design, which is a limiting factor. To offset this, resources are reused whenever possible. In particular, the Montgomery multipliers used to implement the Montgomery exponentiators can also be used whenever single modular multiplications are required, rather than instantiating separate Montgomery multipliers for this purpose.

\subsection{Controller Module Design} \label{subsec:controller}

\begin{algorithm}[t]
\begin{algorithmic}[1]
    \Parameters
        \Desc{$N$}{Paillier public key}
        \Desc{$R$}{Montgomery radix}
        \Desc{$n'$}{Number of bits in mapping from fixed point numbers $\mathbb{Q}(n, m)$ to integers $\mathbb{Z}_{2^{n'}}$}
        \Desc{$T$}{Controller reset period}
        \Desc{$\hat{A}$}{Controller matrix}
        \Desc{$\hat{B}[k]$}{Controller matrix}
    \EndParameters
    \Inputs
        \Desc{$\check{s}$}{Encrypted setpoints in Montgomery form $\check{s}_1, ..., \check{s}_{n_y}$}
        \Desc{$\check{x}$}{Encrypted controller state in Montgomery form $\check{x}_1, ..., \check{x}_{n_x}$}
    \EndInputs
    \Outputs
        \Desc{$\check{x}'$}{Encrypted controller state in Montgomery form $\check{x}'_1, ..., \check{x}'_{n_x}$}
    \EndOutputs
    
    \Function{UpdateState}{$\check{x}, \check{e}, k$}
        \If{$k+1 \mod T = 0$} \Comment{Controller reset}
            \For{$i = 1, ..., n_x$}
                \State $\check{x}'_i \leftarrow R \mod N^2$ \Comment{Encrypted value of 0, in Montgomery form}
            \EndFor
        \Else
            \For{$i = 1, ..., n_y$} \Comment{Generate encrypted error values}
                \State $\check{e}_i \leftarrow \textproc{MontMult}[M=N^2](\textproc{MontExp}(\check{y}, 2^{n'} - 1), \check{s})$
            \EndFor        
            \For{$i = 1, ..., n_x$} \Comment{Generate encrypted scalar products}
                \For{$j = 1, ..., n_x$}
                    \State $\mathrm{var}_{ij} \leftarrow \textproc{MontExp}(\check{x}'_j, \hat{A}_{ij})$
                \EndFor
                \For{$j = 1, ..., n_y$}
                    \State $\mathrm{var}_{i(j+n_x)} \leftarrow \textproc{MontExp}(\check{e}, \hat{B}[k]_{ij})$
                \EndFor
            \EndFor
            \For{$i = 1, ..., n_x$} \Comment{Homomorphically sum up encrypted scalar products}
                \For{$j = 2, ..., n_x + n_y$}
                    \State $\mathrm{var}_{i1} \leftarrow \textproc{MontMult}[M=N^2](\mathrm{var}_{i1}, \mathrm{var}_{ij})$
                \EndFor
                \State $\check{x}'_i \leftarrow \mathrm{var}_{i1}$
            \EndFor
        \EndIf
        \State \Return $\check{x}'$
    \EndFunction
\end{algorithmic}
\caption{Updating state of the dynamic controller using the Montgomery multiplication and the Montgomery exponentiation.}
\label{alg:updatestate}
\end{algorithm}

Consider dynamic controllers in \eqref{eq:dynamichom}. There are computations for incorporating the state of the controller into the generated control inputs, described in Algorithm~\ref{alg:dynamiccontrol}, and for updating the state of the controller, described in Algorithm~\ref{alg:updatestate}. The update of the controller state can be performed independently of the generation of the control inputs. Figure~\ref{fig:encryption_diagram} illustrates the block diagram for a possible realization of Algorithms~\ref{alg:dynamiccontrol} and~\ref{alg:updatestate}. Because calculating the control inputs are independent of each other, individual computations can all be performed in parallel using $n_u$ copies of the multiplier and exponentiator. These parallelizations allow physical systems with more inputs and outputs to be controlled, without increasing the time required to perform the encryptions and decryptions. However, as a trade-off more hardware resources are required, and so in resource limited scenarios, these computations can be performed sequentially if a longer sampling period is acceptable. In the case that the computations all be performed sequentially, the controller would require only one Montgomery exponentiator module. The matrix multiplications for updating the state of the controller can also be parallelized for each row by utilizing $n_x$ copies of the multiplier and exponentiator. The modular exponentiations can also be performed in parallel, and the results are multiplied together afterwards in a binary tree structure with a latency of $\ceil{\log_2(n_x+n_y)}$ times the latency of the Montgomery multiplication.

\begin{algorithm}[t]
\begin{algorithmic}[1]
    \Parameters
        \Desc{$N$}{Paillier public key}
        \Desc{$R$}{Montgomery radix}
    \EndParameters
    \Inputs
        \Desc{$\hat{y}$}{System outputs $\hat{y}_1, ..., \hat{y}_{n_y}$}
        \Desc{$z$}{Values $z_1, ..., z_{n_y}$ where $z_i = r_i^N \mod N^2$}
    \EndInputs
    \Outputs
        \Desc{$\check{y}$}{Encrypted system outputs in Montgomery form $\check{y}_1, ..., \check{y}_{n_y}$}
    \EndOutputs
    \Function{Encrypt}{$\hat{y}, z$}
        \For{$i = 1, ..., n_y$}
            \State $\mathrm{var1} \leftarrow \textproc{MontMult}[M=N^2](NR \mod N^2, \hat{y}_i)$
            \State $\mathrm{var2} \leftarrow \textproc{MontMult}[M=N^2](\mathrm{var1} + 1, R^2 \mod N^2)$
            \State $\check{y}_i \leftarrow \textproc{MontMult}[M=N^2](z_i, \mathrm{var2})$
        \EndFor
        \State \Return $\check{y}$
    \EndFunction
\end{algorithmic}

\caption{Encryption of the system outputs in the plant interface (or of the setpoints elsewhere) using the Montgomery multiplication and the Montgomery exponentiation.}
\label{alg:encryptsensors}
\end{algorithm}

\subsection{Plant Interface Module Design} \label{subsec:plantinterface}

The plant interface's role in the encrypted control system is to encrypt the plant outputs and decrypt the control inputs. There is no requirement for a single plant interface that performs both encryptions and decryptions, as these functionalities can be separated into distinct modules if the actuators and the sensors are physically apart. However, a single plant interface module allows for the reuse of hardware resources for both encryption and decryption, reducing the hardware cost of the system. 

\begin{algorithm}[t]
\begin{algorithmic}[1]
    \Parameters
        \Desc{$N$}{Paillier public key}
    \EndParameters
    \Inputs
        \Desc{$r$}{Random values $r_1, ..., r_{n_y}$}
    \EndInputs
    \Outputs
        \Desc{$z$}{Values $z_1, ..., z_{n_y}$ where $z_i = r_i^N \mod N^2$}
    \EndOutputs
    
    \Function{CalculateRandom}{$r$}
        \For{$i = 1, ..., n_y$}
            \State $z_i \leftarrow \textproc{MontExp}(r_i, N)$
        \EndFor
        \State \Return $z$
    \EndFunction
\end{algorithmic}
\caption{Computing $r^N\mod N^2$ in the plant interface using the Montgomery multiplication and the Montgomery exponentiation.}
\label{alg:random}
\end{algorithm}

\begin{algorithm}[t]
\begin{algorithmic}[1]
    \Parameters
        \Desc{$N$}{Paillier public key}
        \Desc{$R$}{Montgomery radix}
        \Desc{$n'$}{Number of bits in mapping from fixed point numbers $\mathbb{Q}(n, m)$ to integers $\mathbb{Z}_{2^{n'}}$}
        \Desc{$\mu$}{Part of Paillier private key}
        \Desc{$\lambda$}{Part of Paillier private key}
        \Desc{$N^{-1}$}{Value $\in \mathbb{Z}_{N^2+2}$ such that $NN^{-1} \mod (N^2 + 2) = 1$}
    \EndParameters
    \Inputs
        \Desc{$\check{u}$}{Encrypted control inputs in Montgomery form $\check{u}_1, ..., \check{u}_{n_u}$}
    \EndInputs
    \Outputs
        \Desc{$\hat{u}$}{Control inputs $\hat{u}_1, ..., \hat{u}_{n_u}$}
    \EndOutputs
    
    \Function{Decrypt}{$\check{u}$}
        \For{$i = 1, ..., n_u$}
            \State $temp \leftarrow \textproc{MontExp}(\check{u}_i, \lambda)$
            \State $temp \leftarrow \textproc{MontMult}[M=N^2](temp, 1)$
            \State $temp \leftarrow \textproc{MontMult}[M=N^2+2](temp - 1, N^{-1}R^2 \mod (N^2+2))$
            \State $temp \leftarrow \textproc{MontMult}[M=N^2+2](temp, 1)$
            \State $temp \leftarrow \textproc{MontMult}[M=N](temp, \mu R^2 \mod N)$
            \State $\hat{u}_i \leftarrow \textproc{MontMult}[M=N](temp, 1) \mod 2^{n'}$
        \EndFor
        \State \Return $\hat{u}$
    \EndFunction
\end{algorithmic}
\caption{Decryption of the control inputs in the  plant interface using the Montgomery multiplication and the Montgomery exponentiation.}
\label{alg:decrypt}
\end{algorithm}

Paillier encryption algorithm in Algorithm~\ref{alg:encryptsensors} requires values for $r^N \mod N^2$ as inputs, which is independent of the plaintext being encrypted. The steps required for generating $r^N \mod N^2$ are described in Algorithm~\ref{alg:random}. A block diagram similar to Figure~\ref{fig:encryption_diagram} can be employed for parallel realization of the steps in Algorithm~\ref{alg:random}. Note that it is possible to generate the value of $r^N$ needed to encrypt the next system output sample in
parallel with the controller computations involving the encryption of the current sample. This parallelization between the plant interface and controller decreases the time required for completing the necessary tasks within a sampling period without utilizing extra resources. 

There are various approaches for generating cryptographically secure random or pseudo-random values for $r$. Random methods involve sampling a noise source, such as oscillator jitter; examples can be found in \cite{random_osc, random_meta, random_agnostic}. Pseudo-random methods are algorithms that generate numbers from an initial seed, which should be generated from a random method; examples can be found in \cite{random_BBS, random_BM}. Depending on the method used, the generator can be implemented on the FPGA, or external to it. The generated random numbers are used as the input to Algorithm \ref{alg:random}, which first converts them to the Montgomery form, in order to compute $r^N$. Note that, for larger encryption key lengths, checking that $\gcd(r, N) = 1$ is not required, as the probability that this is not the case is negligible. We also do not need to convert random numbers to Montgomery form before performing Montgomery exponentiation. Assume that we are given a uniformly distributed random number $r$ in $\mathbb{Z}_{N}$. With $r'=(rR^{-1})\mod N^2$, where $R$ is the Montgomery radix, it can be seen $r'\mod N=(rR^{-1})\mod N$ is also uniformly distributed random number in $\mathbb{Z}_{N}$ because $r\mapsto (rR^{-1})\mod N$ is bijective ($R$ and $N$ are coprime). Further, $(r'\mod N)^N\mod N^2=(r')^N\mod N^2$ because $(\alpha + kN)^N \mod N^2 = \alpha^N \mod N^2$ for any $\alpha,k\in\mathbb{Z}$. That is, the Montgomery form of $r'$ is in fact $r$. Therefore, using the Montgomery exponentiation algorithm without first converting to Montgomery form, we can compute $r^N\mod N^2$ which is equal to $(r'\mod N)^NR\mod N^2$.

The tasks performed by the plant interface are described in Algorithms~\ref{alg:encryptsensors},~\ref{alg:random}, and~\ref{alg:decrypt}, expressed as a collection of the Montgomery exponentiations and the Montgomery multiplications. The inputs to all of these Montgomery operations are either constants (as the algorithm parameters do not change within any given implementation), algorithm inputs, or the result of the previous operations. Every loop in Algorithms~\ref{alg:encryptsensors},~\ref{alg:random}, and~\ref{alg:decrypt} can be parallelized, as the iterations are independent of each other. For example, the encryptions of plaintexts are independent of each other, so individual encryptions can all be performed in parallel. The same applies to the calculation of values for $r^N \mod N^2$, and to decryptions of the ciphertexts.  If on the other hand the plant interface is fully parallelized, then it would require $\max(n_y, n_u)$ Montgomery exponentiators, as the maximum number of encryptions or decryptions to be performed in parallel depends on whether there are more system outputs to encrypt or more control inputs to decrypt.

\section{Experiment} \label{sec:experiments}

\begin{figure}[t]
    \centering
    \includegraphics[width=0.5\linewidth]{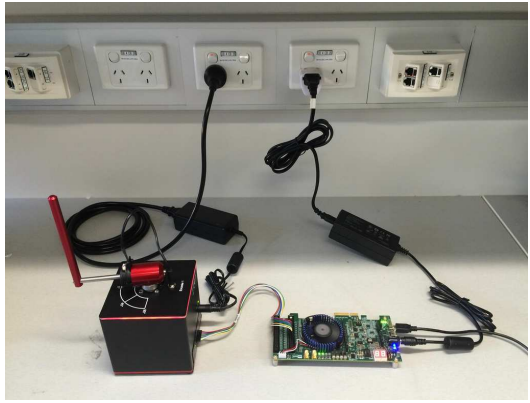}
    \caption{Inverted pendulum balance control experimental setup.}
    \label{fig:setup}
\end{figure}

To demonstrate the system, we have implemented encrypted balance control of an inverted pendulum using our plant interface and controller digital designs on an FPGA. Inverted pendulum systems are unstable and require a dynamic controller to be robustly stabilized. We use the Quanser QUBE-Servo 2 as the plant and the Terasic C5P Development Board (equipped with the Cyclone V GX 5CGXFC9D6F27C7 FPGA) to implement the plant interface and the encrypted controller. The setup is shown in Figure~\ref{fig:setup}.

\begin{figure}
\centering
\begin{tikzpicture}
\node[] at (0,0) {
    \psfrag{x1}[cc][][.8][0]{times (s)}
    \psfrag{x2}[cc][][.8][0]{times (s)}
    \psfrag{x3}[cc][][.8][0]{times (s)}
    \psfrag{v1}[cc][][.8][0]{\raisebox{10pt}{$\theta(^o)$}}
    \psfrag{v2}[cc][][.8][0]{\raisebox{10pt}{$\alpha(^o)$}}
    \psfrag{v3}[cc][][.8][0]{\raisebox{10pt}{$u$ (duty cycle)}}
    \psfrag{y1}[cc][][.7][0]{0}
    \psfrag{y2}[cc][][.7][0]{20}
    \psfrag{y3}[cc][][.7][0]{40}
    \psfrag{z1}[cc][][.7][0]{176}
    \psfrag{z2}[cc][][.7][0]{178}
    \psfrag{z3}[cc][][.7][0]{180}
    \psfrag{z4}[cc][][.7][0]{182}
    \psfrag{w1}[cc][][.7][0]{-0.2}
    \psfrag{w2}[cc][][.7][0]{0}
    \psfrag{w3}[cc][][.7][0]{0.2}
    \psfrag{0}[cc][][.7][0]{0}
    \psfrag{1}[cc][][.7][0]{1}
    \psfrag{2}[cc][][.7][0]{2}
    \psfrag{3}[cc][][.7][0]{3}
    \psfrag{4}[cc][][.7][0]{4}
    \includegraphics[width=0.85\linewidth]{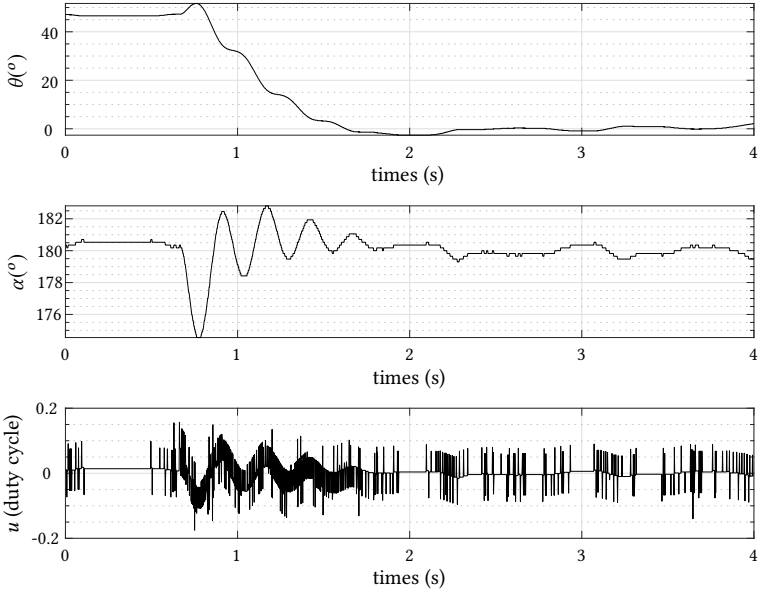}
};
\end{tikzpicture}
\vspace{-.3in}
\caption{The inverted pendulum system stabilizing to its setpoint. Note that the control input duty cycle is signed to specify the direction of rotation for the motor.}
\label{fig:converge}
\end{figure}

We use the following dynamic controller with a control sampling frequency of $500\,\mathrm{Hz}$ to stabilize the inverted pendulum:
\begin{subequations}
\begin{align}
    \mathcal{C} :\quad
        x[k+1] &= \begin{bmatrix}
        0_{3\times 3} & 0_{3\times 1} \\
        \displaystyle\frac{125\pi}{3072} \begin{bmatrix} 
        500 & 0 & 625 \end{bmatrix} & 0
        \end{bmatrix}
        x[k]+\begin{bmatrix}
        I_{3\times 3} \\
        0_{1\times 3}
        \end{bmatrix}(s[k] - y[k])
        \\
        u[k] &= \begin{bmatrix}
        \displaystyle\frac{125\pi}{3072} \begin{bmatrix} 
        -500 & -2 & -655\end{bmatrix} & 1 \end{bmatrix} x[k],
        \\
        s[k] &= \begin{bmatrix} 0 \\ \theta _s[k] \\ 1024 \end{bmatrix},\quad 
        y[k] = \begin{bmatrix} \theta[k] \\ \theta[k] \\ \alpha[k] \end{bmatrix},
\label{eq:qubecontroller}
\end{align}
\end{subequations}
where $\theta[k]$ is the measured rotational arm angle, $\theta _s[k]$ is the rotational arm angle setpoint, $\alpha[k]$ is the measured pendulum angle, all in encoder counts (with 2048 encoder counts measured per revolution), $0_{n\times m}$ is a matrix of zeros with $n$ rows and $m$ columns, and $I_{n\times n}$ is an identity matrix of size $n$. The resulting control input $u$ is a number between $-999$ and $999$, representing a duty cycle and direction. We implement this controller using $n' = 32$ bits, $m = 7$ bits, and an encryption key length of 256 bits. In Section~\ref{sec:secure}, as there were no assumptions on the integer or fractional nature of the parameters, all parameters were multiplied by $2^m$ to generate equivalent integer numbers. However, in this experiment, the sensor measurements and the $C$ matrix are already integers, so we use the following substitutions in our encrypted system:
\begin{subequations}
\begin{align}
\hat{s}_i[k]&=\bar{s}_i[k]\mod 2^{32}\\
\hat{y}_i[k]&=\bar{y}_i[k]\mod 2^{32}\\
\hat{B}_{ij}&=2^7 \bar{B}_{ij}\mod 2^{32}\\
\hat{C}_{ij}&=\bar{C}_{ij}\mod 2^{32}\\
\hat{x}_i[k]&=2^7 \bar{x}_i[k] \mod 2^{32}\\
\hat{u}_i[k]&=2^7 \bar{u}_i[k] \mod 2^{32}
\end{align}
\end{subequations}

Since there is no state evolution (i.e., the state is a simple two steps delay to calculate velocities from position measurements by first order
difference), a resetting the controller state is not required. Rounding and clamping of the generated control input is performed externally from the plant interface and controller.

\begin{figure}[t]
\centering
\begin{tikzpicture}
\node[] at (0,0) {
    \psfrag{x1}[cc][][.8][0]{times (s)}
    \psfrag{x2}[cc][][.8][0]{times (s)}
    \psfrag{x3}[cc][][.8][0]{times (s)}
    \psfrag{v1}[cc][][.8][0]{$\theta(^o)$}
    \psfrag{v2}[cc][][.8][0]{$\alpha(^o)$}
    \psfrag{v3}[cc][][.8][0]{$u$ (duty cycle)}
    \psfrag{y1}[cc][][.7][0]{0}
    \psfrag{y2}[cc][][.7][0]{20}
    \psfrag{y3}[cc][][.7][0]{40}
    \psfrag{z1}[cc][][.7][45]{160}
    \psfrag{z2}[cc][][.7][45]{180}
    \psfrag{z3}[cc][][.7][45]{200}
    \psfrag{w1}[cc][][.7][0]{-0.5}
    \psfrag{w2}[cc][][.7][0]{0}
    \psfrag{w3}[cc][][.7][0]{0.5}
    \psfrag{0}[cc][][.7][0]{0}
    \psfrag{1}[cc][][.7][0]{1}
    \psfrag{2}[cc][][.7][0]{2}
    \psfrag{3}[cc][][.7][0]{3}
    \psfrag{4}[cc][][.7][0]{4}
    \includegraphics[width=0.85\linewidth]{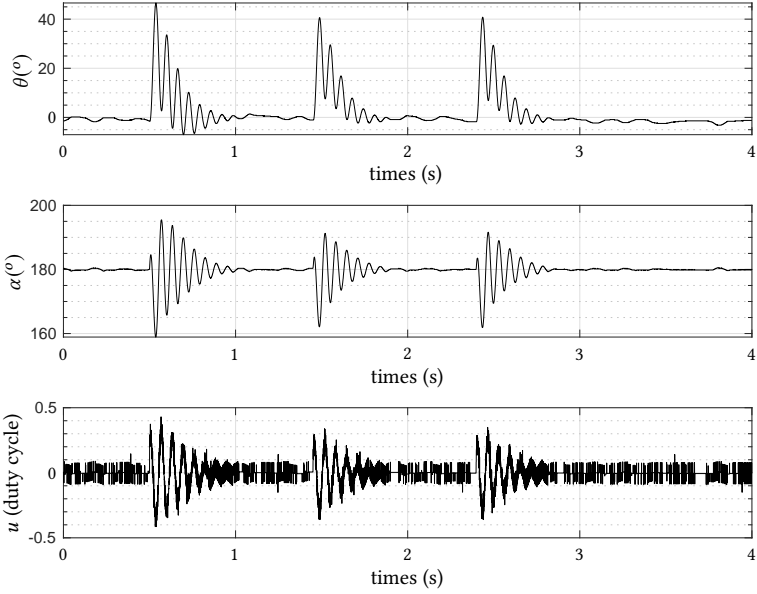}
};
\end{tikzpicture}
\vspace{-.3in}
\caption{The inverted pendulum system with disturbances introduced at the tip of the pendulum. Note that the control input duty cycle is signed to specify the direction of rotation for the motor.}
\label{fig:perturb_tip}
\end{figure}

We utilize the Montgomery multiplier design in Algorithm~\ref{alg:cios2}, which has an embedded multiplier usage that scales linearly with encryption key length. We run two Montgomery multipliers in parallel in each Montgomery exponentiator, and run a single Montgomery exponentiator in the plant interface and controller modules. We neglect the generation of random numbers, but still calculate a number to the power $N$ in each control sampling period. We also neglect instantiating a separate module to encrypt setpoints, and instead encrypt setpoint in the controller, without the use of random numbers. Neither of these simplifications affect the synthesis, timing, or synthesis of the digital design, as the random number generation can be done outside of the digital engine using commercially available integrated circuits for random number generation, and the encryption of setpoints with random numbers can occur in parallel with the encryption of system outputs, thus not extending the minimum control sampling period. Importantly, on the FPGA we have distinct plant interface and controller modules and use an abstracted network to communicate encrypted data between them.

\begin{figure}[t]
\centering
\begin{tikzpicture}
\node[] at (0,0) {
    \psfrag{z1}[cc][][.7][0]{\raisebox{5pt}{$10^{-5}$\quad}}
    \psfrag{z2}[cc][][.7][0]{$10^{-4}$\quad }
    \psfrag{z3}[cc][][.7][0]{$10^{-3}$\quad }
    \psfrag{z4}[cc][][.7][0]{$10^{-2}$\quad }
    \psfrag{x1}[cc][][.7][0]{64}
    \psfrag{x2}[cc][][.7][0]{128}
    \psfrag{x3}[cc][][.7][0]{256}
    \psfrag{x4}[cc][][.7][0]{512}
    \includegraphics[width=0.5\linewidth]{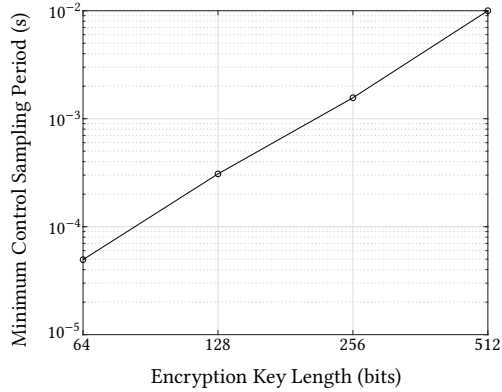}
};
\node[] at (0,-2.6) {\footnotesize Encryption Key Length (bits)};
\node[rotate=90] at (-3.4,0) {\footnotesize Minimum Control Sampling Period (s)};
\end{tikzpicture}
\caption{Graph depicting how the minimum control sampling period increases with greater security.}
\label{fig:time}
\end{figure}

\begin{figure}
\centering
\begin{tikzpicture}
\node[] at (0,0) {
    \psfrag{Encryption Key Length \(bits\)}[cc][][.8][0]{\raisebox{-10pt}{Encryption Key Length (bits)}}
    \psfrag{ALMs used}[cc][][.8][0]{\raisebox{+10pt}{{\color{matlab_blue}ALMs used}}}
    \psfrag{64}[cc][][.7][0]{64}
    \psfrag{128}[cc][][.7][0]{128}
    \psfrag{256}[cc][][.7][0]{256}
    \psfrag{512}[cc][][.7][0]{512}
    \psfrag{x1}[cc][][.7][0]{{\color{matlab_blue}10}}
    \psfrag{x2}[cc][][.7][0]{{\color{matlab_red}10}}
    \psfrag{y1}[cc][][.5][0]{{\color{matlab_blue}3}}
    \psfrag{y2}[cc][][.5][0]{{\color{matlab_blue}4}}
    \psfrag{y3}[cc][][.5][0]{{\color{matlab_blue}5}}
    \psfrag{z1}[cc][][.5][0]{{\color{matlab_red}1}}
    \psfrag{z2}[cc][][.5][0]{{\color{matlab_red}2}}
    \psfrag{z3}[cc][][.5][0]{{\color{matlab_red}3}}
    \psfrag{DSP Blocks used}[cc][][.8][0]{\raisebox{-10pt}{{\color{matlab_red}DSP Blocks used}}}
    \psfrag{ALM usage}[cc][][.5][0]{ALM usage}
    \psfrag{DSP Block usage}[cc][][.5][0]{DSP Block usage}
    \includegraphics[width=0.5\linewidth]{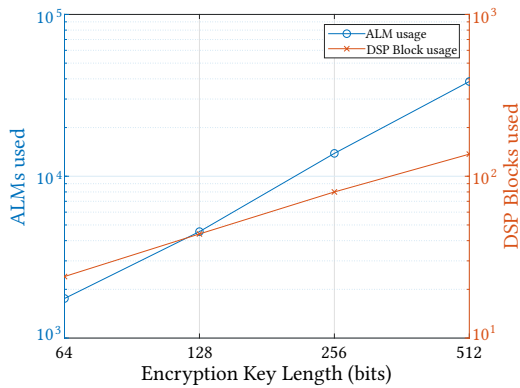}
};
\end{tikzpicture}
\caption{Graph depicting how the usage of hardware resources in the plant interface increases with greater security.}
\label{fig:plant_size}
\end{figure}

Figure \ref{fig:plant_size} shows the hardware resource usage of the plant interface module as the encryption key length increases, for our implementation. Figure \ref{fig:time} shows the minimum control sampling period as the encryption key length increases from 64\,bits to 512\,bits, which affects the speed with which physical systems can be controlled. For the key length of 512\,bits, the sampling time of system is 10\,ms. Implementations using other Montgomery multiplier architectures can potentially result in completely different hardware resource usages and speeds. Such issue are the topic of future work.

Figure \ref{fig:converge} shows the system behaviour converging to its setpoint. Figure \ref{fig:perturb_tip} shows the system behaviour when disturbances are introduced at the tip of the pendulum. Evidently, the controller successfully attenuates large disturbances (of peak magnitude of twenty degrees).

In the experiments, we found that the latency of the plant interface determines the maximum control sampling frequency. This is due to Montgomery exponentiations with the large exponents $N$ and $\lambda$, which require more Montgomery multiplications compared to the Montgomery exponentiations in the controller, where the exponents are shorter. If a larger control sampling frequency is required, then the plant interface digital design could make use of the Chinese Remainder Theorem~\cite{CRT} to reduce the size of the modulus in Montgomery exponentiations, speeding up each calculation.

The hardware description language (HDL) code used for synthesizing the encryption, controller, and decryption in the experiment can be found at \url{https://github.com/availn/EncryptedControl}. A video of the experiment can also be found at \url{https://youtu.be/ATM0tcecst0}.

\section{Conclusions and Future Work} \label{sec:discussions}
We presented an experimental setup to demonstrate a powerful framework for encrypted dynamic control of unstable systems using digital designs on FPGAs with deterministic latency. The framework is scalable and can be applied to large-scale cyber-physical systems. Future work includes investigation of methods for speeding up the computations and studying the effect of uncertain communication systems on the performance of the system.

\begin{acks}

\end{acks}

\bibliographystyle{ACM-Reference-Format}
\bibliography{sample-bibliography}

\end{document}